\documentstyle[12pt]{article}

\input epsf.sty

\newcommand{\insertfig}[2]{\mbox{\epsfxsize=#1cm \epsfbox{#2.eps}}}
\newcommand{\ft}[2]{{\textstyle\frac{#1}{#2}}}

\newcommand{\g}{{\sl g}}

\newcommand{\cD}{{\cal D}}
\newcommand{\cK}{{\cal K}}

\newcommand{\cI}{{\cal I}}
\newcommand{\cO}{{\cal O}}
\newcommand{\cH}{{\cal H}}
\newcommand{\cM}{{\cal M}}

\newcommand{\cN}{{\cal N}}

\newcommand{\cQ}{{\cal Q}}

\newcommand{\cE}{{\cal E}}

\newcommand{\cT}{{\cal T}}

\newcommand{\OO}{\mathop{\otimes}}

\font\cmss=cmss12 
\def\1{\hbox{{1}\kern-.25em\hbox{l}}}
\def\bfZ{\relax{\hbox{\cmss Z\kern-.4em Z}}}

\begin{document}

\begin{titlepage}

\centerline{\large \bf Off-forward gluonometry.}

\vspace{15mm}

\centerline{\bf A.V. Belitsky$^a$, D. M\"uller$^b$}

\vspace{10mm}

\centerline{\it $^a$C.N.\ Yang Institute for Theoretical Physics}
\centerline{\it State University of New York at Stony Brook}
\centerline{\it NY 11794-3840, Stony Brook, USA}

\vspace{5mm}

\centerline{\it $^b$Institut f\"ur Theoretische Physik,
                Universit\"at Regensburg}
\centerline{\it D-93040 Regensburg, Germany}

\vspace{15mm}

\centerline{\bf Abstract}

\vspace{0.5cm}

We explore the deeply virtual Compton scattering process at NLO with the
emphasis on the measurement of the tensor gluon skewed distributions. We
recalculate the one-loop coefficient function and give two-loop results
for anomalous dimensions and exclusive evolution kernels required for
studying the logarithmic scaling violation. We evaluate  weighted cross
sections which give direct access to these functions on experiment.

\vspace{7.5cm}


\vspace{0.5cm}


\end{titlepage}

\section{Introduction.}

The studies of the gluonic content of hadrons are complicated by the fact
that there exists no probes which couples directly to it. The only way for
electroweak bosons to fuse with coloured gluons is via quark loops.
Obviously this is an effect suppressed in the QCD coupling constant on the
background of dominating tree level quark contributions. Gluon distributions
mix logarithmically with quark ones and although they can affect the shape
of the latter in a significant way it is still extremely complicated to
disentangle them in perturbative evolution. An opportunity to measure gluons
directly could appear provided there is a selection rule which forbids
quarks to enter as a leading effect in the cross section.

At leading twist-two level there are three non-local light-ray operators
bilinear in gluon fields. E.g.\ in the light-cone gauge where only
transversely polarized vector fields do propagate, we have the following
projections for the Lorentz structure of the composite operator $B_{\mu}
(x_1) B_{\nu} (x_2) g^\perp_{\mu\mu^\prime} g^\perp_{\nu\nu^\prime}$
\begin{eqnarray}
\label{Decomp}
g^\perp_{\mu\mu'} g^\perp_{\nu\nu'}
= \ft12 g^\perp_{\mu\nu} g^\perp_{\mu'\nu'}
+
\ft12 \epsilon^\perp_{\mu\nu} \epsilon^\perp_{\mu'\nu'}
+ \tau^\perp_{\mu\nu;\rho\sigma} \tau^\perp_{\mu'\nu';\rho\sigma} ,
\end{eqnarray}
where $g^\perp_{\mu\nu} = g_{\mu\nu} - n_\mu n^\star_\nu - n^\star_\mu
n_\nu$, $\epsilon^\perp_{\mu\nu} = \epsilon_{\mu\nu\rho\sigma} n^\star_\rho
n_\sigma$ and $\tau^\perp_{\mu\nu;\rho\sigma} = \frac{1}{2} \left(
g^\perp_{\mu\rho} g^\perp_{\nu\sigma}\right. + g^\perp_{\mu\sigma}
g^\perp_{\nu\rho} - \left. g^\perp_{\mu\nu} g^\perp_{\rho\sigma} \right)$.
Here $n_\mu$ and $n^\star_\nu$ are two light-cone vectors such that $n^2 =
n^{\star\, 2} = 0$ and $n n^\star = 1$. Eq.\ (\ref{Decomp}) corresponds to
the Clebsch-Gordon decomposition of the direct product of the two vector
representations of the Lorentz group\footnote{Recall that the
representations of the Lorentz group $L_+^\uparrow = SO (3,1) = SO (4, {\bf
C})_{\downarrow R} \approx \left( SL (2, {\bf C}) \otimes SL (2, {\bf C})
\right)_{\downarrow R}$ are labeled by a pair $\left( j_1, j_2 \right)$
which are eigenvalues $j_i(j_i + 1)$ of the $SL (2)$ Casimir operators
$\mbox{\boldmath$\hat J$}^2_i$.} $\left( \ft12, \ft12 \right) \otimes \left(
\ft12, \ft12 \right) = \left( 0, 0 \right) \oplus \left( \left( 1, 0 \right)
\oplus \left( 0, 1 \right) \right) \oplus \left( 1, 1 \right)$. The first
two tensors stand for vector and axial sectors and they contribute to the
conventional deep inelastic scattering (DIS) process on spin-$\ft12$
hadrons. The last one cannot appear there since it requires the photon
helicity to be flipped by two units. However, it definitely can appear in
scattering on spin $J \geq 1$ targets as was emphasized in Ref.\
\cite{JafMan89}. Since these operators belong to the spin-2 representation
$\left( 1, 1 \right)$ of the Lorentz group, they cannot mix with quark
operators. Therefore they can serve as a clean probe of the gluonic content
in hadrons not contaminated by other effects. Recently this issue was
discussed in the context of current fragmentation in DIS \cite{SchSzyTer99}
and in the production of two pions in $\gamma \gamma$ fusion
\cite{KivManPol99}. Nevertheless this operator can appear in the Compton
scattering on the nucleon provided it is sandwiched between states with
different momenta as was pointed out \cite{DieGouPirRal97} and elaborated in 
Ref.\ \cite{JiHoo98}. The corresponding kinematics of the process is known as 
deeply virtual Compton scattering (DVCS) \cite{MueRobGeyDitHor94,Ji97,Rad96}. 
In this note we address the issue of the tensor gluon skewed parton 
distributions (SPD) in DVCS in great detail. We recalculate the one-loop 
coefficient function, and present two-loop anomalous dimensions and exclusive 
evolution kernels for this sector. We give an explicit prediction for 
weighted cross sections which can be used to extract directly the tensor 
gluon SPD from experimental data.

\section{Leading twist amplitude.}

\begin{figure}[t]
\begin{center}
\hspace{1cm}
\mbox{
\begin{picture}(0,60)(100,0)
\put(-60,0){\insertfig{10}{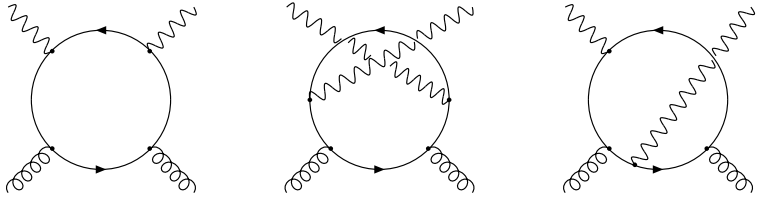}}
\end{picture}
}
\end{center}
\caption{\label{bubble} One-loop diagrams for the coefficient function.}
\end{figure}

The hadronic part of the deeply virtual Compton scattering amplitude
is defined by the off-forward matrix element of the correlator of two
electro-magnetic currents sandwiched between states with unequal momenta
\begin{eqnarray}
\label{EMcurrents}
T_{\mu\nu} (q, P_1, P_2) =
i \int dx e^{i x \cdot q}
\langle P_2 | T j_\mu (x/2) j_\nu (-x/2) | P_1 \rangle ,
\end{eqnarray}
where $q = (q_1 + q_2)/2$ (and the index $\mu$ refers to the outgoing real
photon with momentum $q_2$) and $P_1$ ($P_2$) is the momentum of incoming
(outgoing) nucleon. The leading contribution of the light-ray tensor gluon
operator\footnote{The path-ordered link factor $\Phi [x_2, x_1]$ ensures
gauge invariance.}
\begin{eqnarray}
\label{TransOper}
{^G\!{\cO}^T_{\mu\nu}}
(\kappa_1,\kappa_2)\!
=
G_{+ \rho} (\kappa_2 n) \tau^\perp_{\mu\nu;\rho\sigma}
\Phi [\kappa_2 n, \kappa_1 n] G_{\sigma +} (\kappa_1 n) ,
\end{eqnarray}
whose off-forwards matrix element is parametrized via two SPDs
\cite{JiHoo98}
\begin{eqnarray}
\label{gluonSPD}
&&\!\!\!\!\!\!\!\!\!\!\!G^T_{\mu\nu} (t, \eta, \Delta^2)
\equiv 4 P_+^{-1} \int \frac{d\kappa}{2\pi} e^{i \kappa t P_+}
\langle P_2 |
{^G\!{\cO}^T_{\mu\nu}} (\kappa, -\kappa)
| P_1 \rangle
= H^T_G (t, \eta, \Delta^2)
\frac{\tau^\perp_{\mu\nu;\alpha\beta}}{2 M}
\frac{\Delta_\alpha q_\gamma}{P \cdot q}
\bar U (P_2)
i \sigma_{\gamma\beta} U (P_1) \nonumber\\
&&\qquad\qquad\qquad
+ E^T_G (t, \eta, \Delta^2)
\frac{\tau^\perp_{\mu\nu;\alpha\beta}}{4 M^2} \Delta_\alpha
\bar U (P_2)
\left(
\frac{\Delta_\beta \not\!q}{P \cdot q} - \eta \gamma_\beta
\right)
U (P_1) ,
\end{eqnarray}
into the operator product expansion of currents (\ref{EMcurrents})
appear at one-loop order (see Fig.\ \ref{bubble}). The traceless
symmetric projector $\tau^\perp$ in Eq.\ (\ref{gluonSPD}) possesses
the properties $\tau^\perp_{\mu\nu;\rho\sigma}
\tau^\perp_{\mu\nu;\rho'\sigma'} = \tau^\perp_{\rho\sigma;\rho'\sigma'}$,
$\tau^\perp_{\mu\nu;\rho\sigma} = \tau^\perp_{\rho\sigma;\mu\nu}$,
$\tau^\perp_{\mu\mu;\rho\sigma} = 0$, $\tau^\perp_{\mu\nu;\mu\nu} = 2$.
The kinematical variables used here and below are introduced as $\omega
\equiv \xi^{-1} = - P \cdot q/q^2$ (generalized Bjorken variable),
$\eta = \Delta \cdot q/P \cdot q$ (skewedness), in terms of the vectors
$P = P_1 + P_2$, $q = \ft12 (q_1 + q_2)$ and $\Delta = P_2 - P_1 = q_1
- q_2$.

A simple calculation of one-loop diagrams (see Fig.\ \ref{bubble}) gives us
the following result, with restriction of the reality of the final photon
being relaxed, i.e.\ $q_2^2 \neq 0$,
\begin{equation}
T_{\mu\nu} = - \frac{\alpha_s}{\pi} T_F \sum_{i = 1}^{N_f} Q^2_i
\int_{- 1}^{1} dt \ G^T_{\mu\nu} (t, \eta, \Delta^2) \sigma (t, \eta)
\left\{
1 + \frac{1 - \omega^2 \eta^2}{\omega^2 (t^2 - \eta^2)}
\ln \frac{1 - t^2 \omega^2}{1 - \eta^2 \omega^2}
\right\} ,
\end{equation}
which agrees with Ref.\ \cite{JiHoo98}. The function $\sigma (t, \eta)$
appears from the conversion of gluon field (taken in the light-cone gauge)
into strength tensor, $G_{+\mu} = \partial_+ B_\mu$, and reads, for the
fixing of the residual gauge symmetry  consistent with canonical 
hamiltonian formalism \cite{Bas91},
\begin{eqnarray}
\sigma (t, \eta) &=& \frac{1}{(t - \eta + i0)(t + \eta - i0)} .
\end{eqnarray}

Apart from (suppressed) momentum dependence of the coupling constant an
additional source of scaling violation results from the renormalization
of the composite operator (\ref{TransOper}) which will be discussed in
the next two sections at one- and two-loop order.

\section{Evolution: anomalous dimensions.}

At leading order in the coupling constant the light-ray operator
(\ref{TransOper}) obeys the light-cone position evolution equation
of the form (here and below $\bar y \equiv 1 - y$)
\begin{equation}
\frac{d}{d \ln \mu^2} [ {^G\!{\cO}^T( \kappa_1, \kappa_2 )} ]
= \frac{\alpha_s}{2 \pi}
\int_{0}^{1} dz \int_{0}^{\bar z} dy
{^{GG}\!\cK^T} ( y, z )
[ {^G\!{\cO}^T( \bar y \kappa_1 + y \kappa_2 ,
z \kappa_1 + \bar z \kappa_2 )} ] \nonumber\\
\end{equation}
with the kernel
\begin{equation}
\label{LCPevkertrans}
{^{GG}\!\cK^T} ( y, z ) =
C_A \left\{ y - 2 + \left[\frac{1}{y} \right]_+ \right\} \delta (z)
+ C_A \left\{ z - 2 + \left[\frac{1}{z} \right]_+ \right\} \delta (y)
- \frac{\beta_0}{2} \delta (y) \delta (z) ,
\end{equation}
where $\beta_0 = \frac{4}{3} T_F N_f - \frac{11}{3} C_A$ is the first
coefficient of the QCD $\beta$-function. Fourier transformation of this
result gives the exclusive evolution kernel known before
\cite{BFKL85,JiHoo98,Bel00}. The tree level conformal invariance of
the QCD Lagrangian allows to diagonalize this equation in the basis
spanned by Gegenbauer polynomials\footnote{We drop in what follows
the Lorentz indices on the operators. Here $\partial \!=
\stackrel{\rightarrow}{\partial} \!\!+\!\! \stackrel{\leftarrow}{\partial}$
and $\stackrel{\leftrightarrow}{\cD} = \stackrel{\rightarrow}{\cD} -
\stackrel{\leftarrow}{\cD}$.}
\begin{equation}
{^G\!{\cO}^T_{jl}}
= G_{+ \rho} (i \partial_+)^{l-1} \tau^\perp_{\mu\nu;\rho\sigma}
C^{5/2}_{j - 1}\!
\left( \stackrel{\leftrightarrow}{\cD}_+ / \partial_+ \right)
\!G_{\sigma +} ,
\end{equation}
which form an infinite dimensional representation of the conformal group
in the space of bilinear operators. Therefore,
\begin{equation}
\label{LOevol}
\frac{d}{d \ln \mu^2} [ {^G\!{\cO}^T_{jl}} ]
= - \frac{1}{2} \sum_{k = 1}^{j}
{^{GG}\!\gamma^{T}_{jk}}\, [ {^G\!{\cO}^T_{jl}} ],
\qquad\mbox{with}\qquad
{^{GG}\!\gamma^{T(0)}_j}
= 4 C_A \left( \psi (j + 2) - \psi (1) \right) + \beta_0 ,
\end{equation}
the first term in the expansion ${^{GG}\!\gamma^{T}_{jk}} = \left(
\frac{\alpha_s}{2 \pi} \right) {^{GG}\!\gamma^{T(0)}_j} \delta_{jk}
+ \left( \frac{\alpha_s}{2 \pi} \right)^2 {^{GG}\!\gamma^{T(1)}_{jk}}
+ \cO (\alpha_s^3)$.

\begin{figure}[t]
\vspace{-1cm}
\begin{center}
\mbox{
\begin{picture}(0,130)(225,0)
\put(0,-2){\insertfig{5}{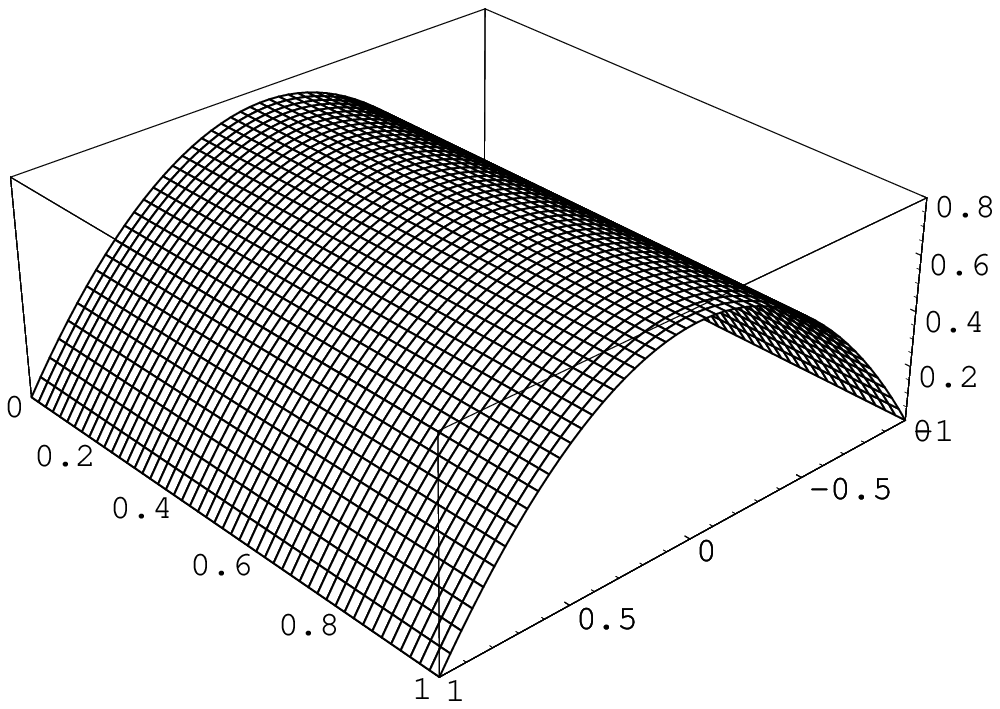}}
\put(150,-2){\insertfig{5}{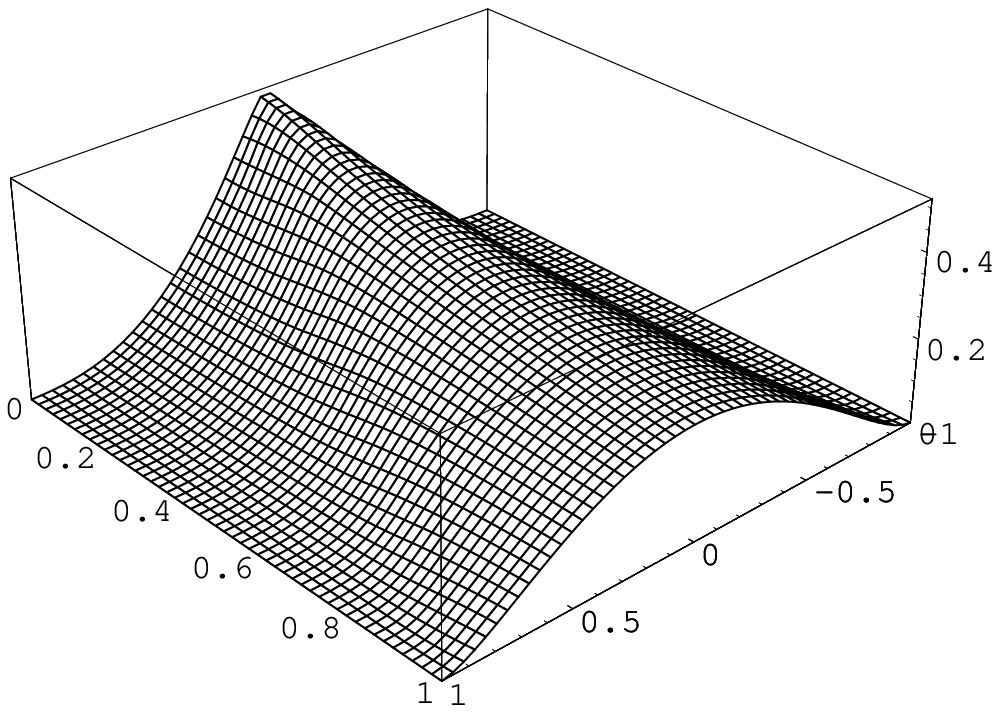}}
\put(300,-2){\insertfig{5}{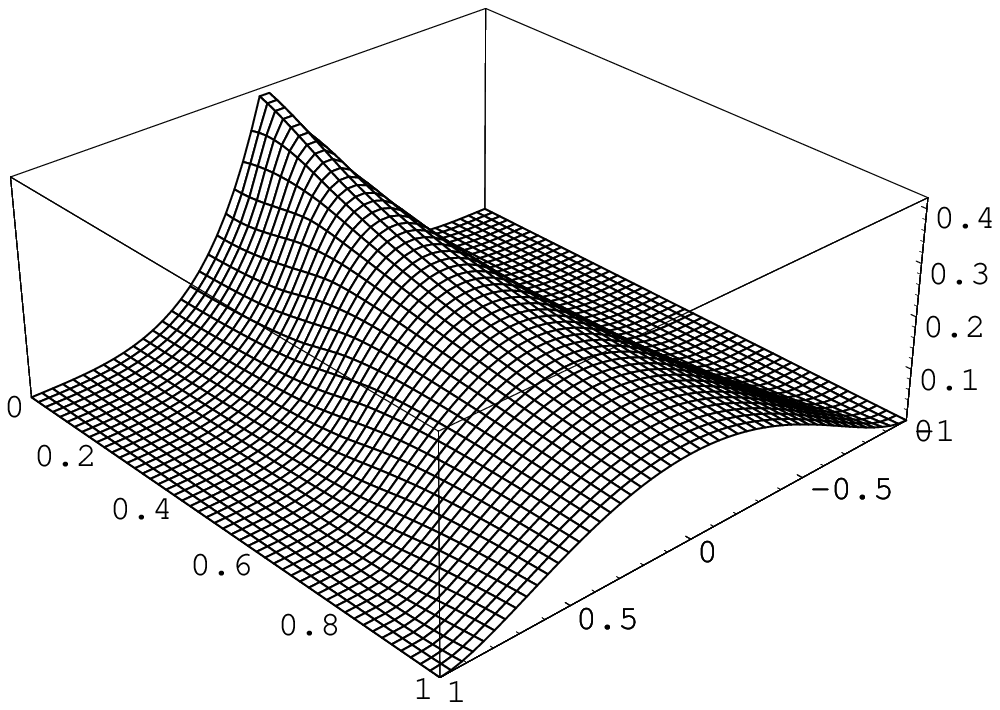}}
\put(83,77){{\small (a)}}
\put(235,77){{\small (b)}}
\put(386,77){{\small (c)}}
\put(10,10){$\eta$}
\put(410,10){$t$}
\end{picture}
}
\end{center}
\caption{\label{evolution} Input SPD in (a) and its evolved shape at $4\
{\rm GeV}^2$ in (b) and at $100\ {\rm GeV}^2$ in (c) for $N_{\rm max}
= 100$.}
\end{figure}

In the momentum fraction space the evolution of the SPD can be done
making use of orthogonal polynomial reconstruction (in the following
Gegenbauer polynomials, $C_j^{5/2}$) of the function from its conformal
moments according to Ref.\ \cite{BelGeyMulSch99}
\begin{equation}
G (t, \eta, Q^2) = \sum_{j = 1}^{N_{\rm max}}
\widetilde C_{j - 1}^{5/2} (t)
\sum_{k = 1}^{j} c_{jk} (\eta)
\left(
\frac{\alpha_s (Q_0^2)}{\alpha_s (Q^2)}
\right)^{{^{GG}\!\gamma_k^{T(0)}}/\beta_0}
\eta^{k - 1} \int_{-1}^{1} dt\, C_{k - 1}^{5/2} (t/\eta) G(t,\eta,Q_0^2) ,
\end{equation}
where formally $N_{\rm max} = \infty$. Here $\widetilde C_{j - 1}^{5/2}
(t) = \frac{9}{2} \frac{(2j + 3)}{(j)_4} \left( 1 - t^2 \right)^2
C_{j - 1}^{5/2} (t)$ are adjoint polynomials and the re-expansion
coefficients $c_{jk} (\eta) = \langle \widetilde C_{k - 1}^{5/2} (t)
| C_{j - 1}^{5/2} (\eta t) \rangle$ are expressed in terms of
hypergeometric function ${_2F_1}(\eta^2)$. The evolution is demonstrated
in Fig.\ \ref{evolution} where we have taken an $\eta$-independent
input\footnote{We factored out the $\Delta^2$ dependence into the gluonic
form factor $F_G (\Delta^2)$ which for phenomenological estimations can
be taken equal $\kappa_T \left( 1 - \Delta^2/ M_{\Lambda}^2 \right)^{-3}$
with $M_{\Lambda} = 2.7 \ {\rm GeV}$ and unknown parameter $\kappa_T$ which
defines the magnitude of proton matrix element of the tensor gluonic
operator.}, $G (t,\eta) = \ft34 (1 - t^2)$, at very low $Q_0^2 = 0.2 \
{\rm GeV}^2$ and evolved it up to $Q^2 = 4 \ {\rm GeV}^2$ (b) and $Q^2
= 100 \ {\rm GeV}^2$ (c).

As a side remark on the solution of the leading order equation for
``transversity" let us note that even in the case when any arbitrary
number of gluons with the same helicity, whose pair-wise interaction
is described by the kernel (\ref{LCPevkertrans}), are exchanged in
the $t$-channel, one can still diagonalize the (LO) multi-particle kernel
since the problem admits a large enough number of conservation laws to
be completely integrable. This can be found by noticing the equivalence
of the anomalous dimensions (\ref{LOevol}) (which depend on the
eigenvalues of $SL (2)$ Casimir operator $\mbox{\boldmath$\hat J$}^2$)
to the Hamiltonian of the exactly solvable one-dimensional $XXX_{s = -3/2}$
spin chain model \cite{Bel00}.

Next we compute the two-loop anomalous dimensions for the tensor gluon
operator. To this end we use our machinery developed in Ref.\
\cite{BelMul98}. Let us give a brief outline of the method. The structure
of anomalous dimensions of conformal operators at two-loop (and higher)
order reads
\begin{equation}
{^{GG}\!\gamma_{jk}^{T(1)}} = {^{GG}\!\gamma_j^{T(1),{\rm D}}} \delta_{jk}
+ {^{GG}\!\gamma_{jk}^{T(1),{\rm ND}}},
\end{equation}
where ${^{GG}\!\gamma_j^{T(1),{\rm D}}}$ are the next-to-leading order (NLO)
forward anomalous dimensions\footnote{We thank W. Vogelsang for providing
us his result for local anomalous dimensions.} \cite{Vog98}
\begin{eqnarray}
{^{GG}\!\gamma_j^{T(1),{\rm D}}} \!\!\!&=&\!\!\!
C_A^2 \left\{ S_1 (j + 1) \left( \frac{134}{9}
- 4 S_2^\prime \left( \frac{j + 1}{2} \right)\right)
- S_3^\prime \left( \frac{j + 1}{2} \right)
+ 8 \tilde{S}(j + 1) - \frac{1}{j (j + 3)} - \frac{16}{3}
\right\}
\nonumber\\
&+&\!\!\! C_A T_F N_f \left\{ \frac{8}{3}
- \frac{40}{9} S_1 (j + 1) - \frac{2}{j ( j + 3)} \right\}
+ C_F T_F N_f \frac{2 (j + 1)(j + 2)}{j (j + 3)} ,
\end{eqnarray}
with
\begin{eqnarray*}
S_\ell (j)
= \sum_{k = 1}^{j} \frac{1}{k^\ell} ,
\qquad
S^\prime_\ell \left( \frac{j}{2} \right)
= 2^\ell \sum_{k = 1}^{j} \frac{\sigma_k}{k^\ell} ,
\qquad
\widetilde{S} (j)
= \sum_{k = 1}^{j} \frac{(-1)^k}{k^2} S_1 (k) ,
\end{eqnarray*}
and where $\sigma_j = \ft12 [1 + (-1)^j]$.
The non-diagonal elements of anomalous dimension matrix of the conformal
operators arise due to one-loop breaking of the conformal symmetry and,
since the tree level conformal invariance leads to diagonal anomalous
dimensions, the one-loop special conformal anomaly generates two-loop
anomalous dimensions. The use of four-dimensional conformal algebra
provides a relation between the anomalies of dilatation (read anomalous
dimensions in question) and special conformal transformations via the
commutator $[\cD, \cK_-] = i \cK_-$ which is applied on the Green function
of elementary fields with conformal operator insertion. To evaluate the
commutator the knowledge of scale and special conformal Ward identities,
with unraveled pattern of symmetry breaking for afore mentioned Green
function, is indispensable. A careful analysis reveals the result
\cite{BelMul98}
\begin{equation}
\label{andimND-GG}
{^{GG}\!\gamma_{jk}^{T(1),{\rm ND}}}
=
\left(
{^{GG}\!\gamma_{j}^{T(0)}} - {^{GG}\!\gamma_{k}^{T(0)}}
\right)
\left\{
d_{jk}
\left(
\beta_0 - {^{GG}\!\gamma_k^{T(0)}}
\right)
+ {^{GG}\!g^T_{jk}}
\right\} ,
\end{equation}
where ${^{GG}\!\gamma_k^{T(0)}}$ are already known LO anomalous dimensions
(\ref{LOevol}), $d_{jk} = - \sigma_{j - k} (2k + 3)/(j - k)\cdot(j + k + 3)$
for $j > k$ and $g_{jk}^T$ appears as
a counterterm required for renormalization of the product of two composite
operators: integrated anomaly $\cO_A^- \equiv \int d^4 x \, 2x_- \cO_A (x)
= \int d^4 x \, x_- Z_3 \left( G_{\mu\nu} \right)^2$ in the trace of
energy-momentum tensor and a conformal operator. The structure of  the
counterterms has been established using the form of counterterms
for differential vertex operator insertions and found to be
\begin{eqnarray}
\label{ren-A}
i[{\cO}_A (x)] [{^G\!\cO^T_{j l}}]
= i[{\cO}_A (x) {^G\!\cO^T_{j l}}]
\!\!\!&-&\!\!\!
\delta^{(d)} (x) \sum_{k = 0}^{j}
\left\{ \hat Z_A \right\}_{jk}
[{^G\!\cO^T_{k l}}]
-
\frac{i}{2} \partial_+ \delta^{(d)} (x) \sum_{k = 0}^{j}
\left\{ \hat Z_A^- \right\}_{jk}
[{^G\!\cO^T_{k l - 1}}]
- \dots
\nonumber\\
\!\!\!&-&\!\!\!
\left(
\g \frac{\partial\ln X}{\partial\g}
- 2 \xi \frac{\partial\ln X}{\partial\xi}
\right)
B_\mu^a (x) \frac{\delta}{\delta B_\mu^a (x)}
[{^G\!\cO^T_{j l}}] ,
\end{eqnarray}
where $X = Z_{\g} Z_3^{1/2}$ is expressed in terms of the renormalization
constants of the gluon wave function $Z_3$ and the coupling $Z_{\g}$.

\begin{figure}[t]
\begin{center}
\hspace{3cm}
\mbox{
\begin{picture}(0,50)(100,0)
\put(0,-5){\insertfig{4}{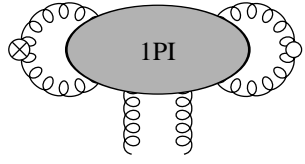}}
\end{picture}
}
\end{center}
\caption{\label{1partirr} 1PI Green function for calculation of the
renormalization constant $Z^-_A$.}
\end{figure}

To determine the unknown renormalization matrix $Z^-_A$ it proves
convenient to work in the light-cone position formalism and calculate
1PI Green function $\langle [{\cO}_A (x)] [{^G\!{\cO}^T} (\kappa_1,
\kappa_2)] B_\mu (x_1) B_\nu (x_2) \rangle_{\rm amp}$ shown in Fig.\
\ref{1partirr} (explicit one-loop graphs can be found in Ref.\
\cite{BelMul98}). We write the divergent part of the operator product
$i [ {\cO}_A^- ] [ {^G\!\cO^T ( \kappa_1, \kappa_2 )} ]$ via the
following relation in leading order of the coupling constant:
\begin{eqnarray}
i [ {\cO}_A^- ] [ {^G\!\cO^T( \kappa_1, \kappa_2 )} ]
&=& \frac{\alpha_s}{2 \pi} \frac{i}{\epsilon}
\int_{0}^{1} dz \int_{0}^{\bar z} dy
\biggl\{
{^{GG}\!\cK_A^-} ( y, z )
[ {^G\!\cO^T( \bar y \kappa_1 + y \kappa_2 ,
z \kappa_1 + \bar z \kappa_2 )} ] \nonumber\\
&+&
{^{GG}\!\widetilde\cK_A^-} ( y, z )
\int d^d x \, 2 x_- B_\mu^b (x)
\frac{\delta}{\delta B_\mu^b (x) }
[ {^G\!{\cO}^T( \bar y \kappa_1 + y \kappa_2 ,
z \kappa_1 + \bar z \kappa_2 )} ]
\biggr\}.
\end{eqnarray}
For the evolution kernel ${^{GG}\!\widetilde\cK_A^-}$ we find immediately
from Eq.\ (\ref{ren-A}) that ${^{GG}\!\widetilde\cK_A^-} ( y, z ) = - i
\ft34 \delta (y) \delta (z)$. Explicit calculation of the amputated Green
function of conformal anomaly and conformal operator gives
\begin{equation}
\label{gluon-ZA}
{^{GG}\!\cK_A^-} ( y, z )
= - i 2 ( \kappa_1 + \kappa_2 ) {^{GG}\!\cK^T} ( y, z )
- i ( \kappa_1 + \kappa_2 ) \beta_0 \delta (y) \delta (z)
+ {^{GG}\!\cK_w^T} ( y, z ),
\end{equation}
where ${^{GG}\!\cK^T} ( y, z )$ is the LO evolution kernel
(\ref{LCPevkertrans}) and
\begin{eqnarray}
\label{KwT}
{^{GG}\!\cK_w^T} ( y, z )
&=& 2 \frac{C_A}{k_{2+}}
\left\{
\left[ \frac{2}{z} \right]_+\delta (y)
-
\left[ \frac{1}{z^2} \right]_+\delta (y)
+ (2y - 1) \delta (z)
\right\} \nonumber\\
&+& 2 \frac{C_A}{k_{1+}}
\left\{
\left[ \frac{2}{y} \right]_+ \delta (z)
-
\left[ \frac{1}{y^2} \right]_+ \delta (z)
+ (2z - 1) \delta (y)
\right\} .
\end{eqnarray}
The evaluation of the conformal moments of (\ref{KwT})
[multiplied by $1/2\cdot (j-k)\cdot (j+k+3)$] gives us
\begin{eqnarray}
{^{GG}\!g_{jk}^T}
\!\!\!&=&\!\!\!
-  2 C_A \sigma_{j - k} \theta_{j - 2,k}
\frac{( 3 + 2k )}{(j - k) (j + k + 3)} \\
&\times&\!\!\!\left\{
2 A_{jk} + ( A_{jk} - \psi (j+2) + \psi(1) )
\left[
\frac{\Gamma (j + 4)\Gamma (k)}{\Gamma (j)\Gamma (k + 4)} - 1
\right]
\right\} , \nonumber
\end{eqnarray}
where we have introduced the matrix $A$ with its elements defined by
\begin{equation}
A_{jk} = \psi\left( \frac{j + k + 4}{2} \right)
- \psi\left( \frac{j - k}{2} \right)
+ 2 \psi ( j - k ) - \psi ( j + 2 ) - \psi(1) ,
\end{equation}
and $\theta_{jk} = 1$ for $j > k$ and zero otherwise. Note that this
result differs from our previous ones for chiral even operators
only by a pure rational function
\begin{equation}
{^{GG}\!g_{jk}^T} = {^{GG}\!g_{jk}^V}
+ 4 C_A \sigma_{j-k} ( 3 + 2k ) \frac{\Gamma (k)}{\Gamma (k + 4)} .
\end{equation}
The solution of two-loop evolution equation is straightforward and
can be found in Ref.\ \cite{BelGeyMulSch99}.

\section{Evolution: two-loop kernel.}

In this section we give our results for the two-loop exclusive evolution
kernel reconstructed from the known non-diagonal anomalous dimensions
found above and two-loop splitting function of Ref.\ \cite{Vog98}. The
structure of ER-BL kernel to $\cO \left( \alpha_s^3 \right)$ accuracy
reads
\begin{equation}
{^{GG} V^T} =
\frac{\alpha_s}{2\pi}
\left\{ \left[ {^{GG} V^{T(0)}} \right]_+
- \ft12 {^{GG} \gamma_1^{T(0)}} \delta (x - y) \right\}
+ \left( \frac{\alpha_s}{ 2 \pi} \right)^2
\left\{ \left[ {^{GG} V^{T(1)}} \right]_+
- \ft12 {^{GG} \gamma_1^{T(1)}} \delta(x - y) \right\} ,
\end{equation}
where the +-prescription is defined according to
\begin{eqnarray}
\left[ V (x, y) \right]_+
= V(x, y) - \delta(x - y) \int_0^1 dz V(z, y) .
\end{eqnarray}
Note that this prescription is not in one-to-one correspondence with the
$+$-prescription in the forward case. The LO kernel reads
\cite{BFKL85,JiHoo98,Bel00}
\begin{eqnarray}
\label{Def-Ker-tra-LO}
{^{GG} V^{T(0)}} = C_A  \theta(y - x) {^{GG}\!f^T} (x, y)
+ \left\{ x \to \bar x \atop y \to \bar y \right\}
\quad\mbox{with}\quad
{^{GG}\!f^T} = \frac{x^2}{y^2} \frac{1}{y - x} .
\end{eqnarray}
To obtain the NLO correction we use the approach described in
\cite{BelFreMue99}. Since the tensor gluon sector is almost analogous
to handle as the vector and axial ones, we only mention the differences
to these cases. The structure of two-loop anomalous dimensions
(\ref{andimND-GG}) implies the following form of the kernel
\begin{eqnarray}
\label{Def-NLO-Ker}
{^{GG} V^{T(1)}}
= - {^{GG} \dot V^T} \otimes
\left( {^{GG} V^{T(0)}} + \frac{\beta_0}{2} \right)
- \left[{^{GG}\!g^T} \OO_{\mbox{'}} {^{GG}V^{T(0)}} \right]
- \frac{ C_A^2}{2} \, {^{GG}\!G^T} + {^{GG}\!D^T} ,
\end{eqnarray}
where the commutator stands for $[ A \, {\displaystyle \OO_{\mbox{'}} } \,
B ] (x, y) = \int_0^1 dz\, \left\{ A (x, z) B (z, y) - B (x, z)
A (z, y) \right\}$. The off-diagonal part is contained in the first two
convolutions on the r.h.s.\ of this equation and they are known exactly. The
conformal moments of the dotted kernel,
\begin{equation}
{^{GG}\dot V^{T(0)}}
= C_A \theta(y - x) \frac{x^2}{y^2} \frac{1}{y - x} \ln\frac{x}{y}
+ \left\{x \to \bar x \atop y \to \bar y \right\} ,
\end{equation}
are proportional to the commutator of the $d$-matrix with the LO
anomalous dimensions. The $g$-matrix corresponds to the kernel
\begin{eqnarray}
{^{GG}\!g^T}
= C_A \theta(y - x)
\left[ 2 \frac{x}{y}
- \frac{\ln\left(1 - \frac{x}{y}\right)}{y - x} \right]
+ \left\{x \to \bar x \atop y \to \bar y \right\} ,
\end{eqnarray}
which contains a part (second term) of the chiral even (odd) case. The
difference is a rational function (first term) which can be restored from
the $GQ$-channel result by applying appropriate convolutions in order to
trade the denominator $1/(j + 1)\cdot(j + 2)$ of the $GQ$ conformal moments
of $x/y$ to  go over into the $GG$ sector\footnote{This can be achieved by 
convolution with the so-called
$c$-kernel \cite{BelFreMue99} in $QQ$ channels and differentiation with
respect to $y$, i.e.\ $d^3 y^2 {\bar y}^2/dy^3$. Finally, we removed a pure
diagonal piece.}, i.e.\  $1/j\cdot(j + 1)\cdot(j + 2)\cdot(j + 3)$.

Now the problem is reduced to the reconstruction of diagonal parts.
The third term reads
\begin{eqnarray}
{^{GG}\!G^T}
\!\!\!&=&\!\!\! \left\{ \theta(y - x)
\left[ {^{GG}\!h^T} (x, y)
+ \Delta {^{GG}\!h^T} (x, y) \right]
+ \theta(y - \bar x)
\left[ {^{GG} \bar h^T} (x, y)
+ \Delta {^{GG}\!h^T} (\bar x, y) \right] \right\} \nonumber\\
&+&\!\!\! \left\{ x \to \bar x \atop y \to \bar y \right\} ,
\end{eqnarray}
and arises from the crossed-ladder diagram and can be obtained by
means of $\cN = 1$ supersymmetry from the one in the quark sector.
The particular contributions are
\begin{eqnarray}
{^{GG}\! h^T}
\!\!\!&=&\!\!\! 2\, {^{GG}\!\bar f^T} \ln{\bar x} \ln{y}
- 2\,{^{GG}\! f^T}
\left[ {\rm Li_2}(x) + {\rm Li_2}(\bar y) \right],
\quad
\Delta{^{GG}\! h^T}
= - \frac{2 x}{y^2 \bar y}
- \frac{2 \bar x}{y^2 \bar y} \ln\bar x
- \frac{2 x}{y \bar y^2} \ln{y} ,
\nonumber\\
{^{GG}\!\bar h^T}
\!\!\!&=&\!\!\! \left( {^{GG}\! f^T - {^{GG}\!\bar f^T}} \right)
\left[ 2 {\rm Li_2}
\left( 1 - \frac{x}{y} \right) + \ln^2 y
\right]
+ 2 \, {^{GG}\! f^T} \left[ {\rm Li_2}(\bar y)- \ln x \ln y \right]
+ 2 \, {^{GG}\!\bar f^T} {\rm Li}_2 (\bar x) .
\nonumber\\
\end{eqnarray}
Here ${\rm Li}_2 (x) = - \int_0^x \ft{dt}{t} \ln (1 - t)$ is the Euler
dilogarithm and we used the following shorthand notation ${^{GG}\!\bar f^T}
= {^{GG}\!f^T} (\bar x, \bar y)$. The remaining diagonal part $D$ has a
simple representation in terms of LO kernels and can be obtained by taking
the forward limit and comparison with the known DGLAP kernel \cite{Vog98}.
Consequent restoration of diagonal ER-BL kernels from the splitting
functions is straightforward and gives
\begin{eqnarray}
\label{Cal-Dia-Rem}
{^{GG}\!D^{T}}
= \!\!\!&-&\!\!\! C_F T_F N_f
\left[ {^{GG}\!v^a} + \frac{2}{3} {^{GG}\!v^c} \right]
+ \beta_0 C_A
\left[ \frac{3}{8} {^{GG}\!v^a} - \frac{5}{6} {^{GG}\!v^b}
+ \frac{1}{4} {^{GG}\!v^c} \right]
\nonumber\\
&+&\!\!\! C_A^2 \left[ \frac{13}{8} {^{GG}\!v^a}
- \frac{11}{6} {^{GG}\!v^b} + \frac{13}{12} {^{GG}\!v^c} \right] .
\end{eqnarray}
Here the $b$ kernel coincides (with colour factor being dropped) with the
LO kernel (\ref{Def-Ker-tra-LO}) and the $a$ and $c$ kernels having the
structure $v^i (x, y) = f^i \theta (y - x) + \bar f^i \theta (x - y)$ are
defined by the functions ${^{GG}\!f^a} = x^2/y^2$ and ${^{GG}\!f^c} = x^2
( 2 \bar x y + y - x )/y^2$. Eqs.\ (\ref{Def-NLO-Ker}-\ref{Cal-Dia-Rem})
is our final result for the NLO corrections to the gluon ``transversity"
evolution kernel.

\section{Cross sections.}

Now we are in a position to study the cross section for electroproduction
of real photon where the gluonic SPD, whose perturbative properties we
studied in detail in the previous sections, can be measured. For DVCS the
skewedness $\eta$ and generalized Bjorken variable $\xi$ are proportional
$\eta = - \xi \left( 1 + \frac{\Delta^2}{2 \cQ^2} \right)^{-1}$. The
differential cross section with unpolarized lepton beam and unpolarized
target, in variables $y = P_1 \cdot q_1 /P_1 \cdot k$, ${\cal Q}^2\equiv
-q_1^2$ and $x \equiv \cQ^2 / (2 P_1 \cdot q_1)$ with $\xi = x \left( 1
+ \frac{\Delta^2}{2 \cQ^2} \right) \left(2 - x + x \frac{\Delta^2}{\cQ^2}
\right)^{-1}$, reads
\begin{eqnarray}
\frac{d\sigma}{dx dy d|\Delta^2| d\phi_r}
=
\frac{\alpha^3  x y } { 8 \pi \, {\cal Q}^2}
\left( 1 + \frac{4 M^2 x}{{\cal Q}^2} \right)^{-1/2}
\left| \frac{\cT}{e^3} \right|^2 .
\end{eqnarray}
Here $\phi_r$ is the azimuthal angle between the lepton and proton
scattering planes in the rest frame of the target. In the following
we will be interested only in the interference term $ |{\cal I}|^2 \equiv
{\cal T}_{\rm BH} {\cal T}_{\rm DVCS}^\ast + {\cal T}_{\rm DVCS}
{\cal T}_{\rm BH}^\ast$ between the DVCS, ${\cal T}_{\rm DVCS}$, and
Bethe-Heitler amplitude ${\cal T}_{\rm BH}$ since it provides a unique
opportunity to extract the real/imaginary part of the gluonic amplitudes
\begin{equation}
\left\{
\begin{array}{c}
\cH^T_G (\xi, \Delta^2) \\
\cE^T_G (\xi, \Delta^2)
\end{array}
\right\}
= - \frac{\alpha_s}{\pi} T_F \sum_{i = 1}^{N_f} Q^2_i
\int_{- 1}^{1} dt \
\sigma (t, - \xi)
\left\{
\begin{array}{c}
H^T_G (t, \xi, \Delta^2) \\
E^T_G (t, \xi, \Delta^2)
\end{array}
\right\} .
\end{equation}
An explicit calculation gives for unpolarized settings
\begin{eqnarray}
\label{Inteference}
\left| \frac{\cI}{e^3} \right|^2
\!\!\!&=&\!\!\! \frac{( \pm 1 )}{2 \cQ^2 \Delta^2}
{\rm Sp} \left\{ \not\!k \left[
\gamma_\gamma (\not\!k - \not\!\!\Delta)^{-1} \gamma_\mu
+ \gamma_\mu (\not\!k^\prime + \not\!\!\Delta)^{-1} \gamma_\gamma
\right] \not\!k^\prime \gamma_\nu \right\} \nonumber\\
&&\qquad\qquad\qquad\qquad\qquad\times
\tau^\perp_{\mu\nu;\alpha\beta} \Delta_\alpha
\left\{
\left( \xi g_{\beta\gamma} + \frac{\Delta_\beta q_\gamma}{P \cdot q} \right)
\cT_1
+ P_\gamma \Delta_\beta \cT_2
\right\} ,
\end{eqnarray}
where
\begin{equation}
\cT_1 = 2 (F_1 + F_2) {\rm Re}
\left( \cH^T_G + \frac{\Delta^2}{4 M^2} \cE^T_G \right) , \qquad
\cT_2 = \frac{1}{2 M^2} {\rm Re}
\left( F_1 \cE^T_G - F_2 \cH^T_G \right) .
\end{equation}
and $+(-)$ sign stands for electron (positron) beam. In consequent evaluation
of this expression we perform an expansion in $1/\cQ^2$ and keep only the
first non-vanishing contribution. We form the charge asymmetry in order to
extract the interference term from the total cross section. Since the double
helicity flip amplitude presently considered is uniquely proportional to
$\cos \left( 3 \phi_r \right)$ (while all other contributions to the
non-polarized cross section $|\cI|^2$ enter with $\cos (n \phi_r)$ and $n
= 1, 2$ \cite{DieGouPirRal97,BelMulNieSch00} one can isolate this
purely gluonic contributions from the effects of quark SPD by forming an
appropriate weighted cross section $\int d \phi_r w (\phi_r) \sigma
(\phi_r)$ \cite{DieGouPirRal97}. Namely, choosing $w (\phi_r) = \cos
\left( 3 \phi_r \right)$ we extract tensor gluon SPD from the unpolarized
cross section,
\begin{eqnarray}
\label{WeightCSunp}
\frac{1}{2\pi} \int_{0}^{2 \pi} d\phi_r
\cos \left( 3 \phi_r \right)
\frac{d^{+} \sigma - d^{-} \sigma}{d\phi_r}
= 16 \sqrt{-\frac{\Delta^2}{\cQ^2}}
\frac{\sqrt{(1 - x)^3 (1 - y)}}{x y\, (2 - x)^2}
\left( 1 - \frac{\Delta^2_{\rm min}}{\Delta^2} \right)^{3/2}
\cT_2 \, d \cM ,
\end{eqnarray}
where $d\cM = \frac{\alpha^3 x y} { 8 \, \pi \, {\cal Q}^2} \left( 1 +
\frac{4 M^2 x}{{\cal Q}^2} \right)^{-1/2} dx dy d|\Delta^2|$ and
$\Delta_{\rm min}^2 = - M^2 x^2/ (1 - x + x M^2/\cQ^2)$. Note that the
correction to the structure function $\cT_2$ is suppressed by $\cO \left(
\sqrt{ - \Delta^2/\cQ^2} \right)$, while the combination $\cT_1$ is down
by $\cO \left( M^2 / \sqrt{- \Delta^2 \cQ^2} \right)$ relative to $M^2
\cT_2$. 

The imaginary part of the amplitudes $\cH^T_G$ and $\cE^T_G$ 
can be accessed by means of single spin asymmetry. 
The polarization of the lepton beam does not induce contribution of gluon
``transversity'' distributions into the cross section in agreement with 
\cite{DieGouPirRal97}. However, once the proton beam is longitudinally
polarized we can access this asymmetry due to the interference term 
since the latter has genuine $\sin \left( 3 \phi_r \right)$ azimuthal angle
dependence. Thus the SPD can be extracted via the following weighted cross
section
\begin{eqnarray}
\label{WeightCSpol}
\frac{1}{2\pi} \int_{0}^{2 \pi} d\phi_r
\sin \left( 3 \phi_r \right)
\frac{d^{+} \sigma_{\rightarrow} - d^{+} \sigma_{\leftarrow}}{d\phi_r}
= 16 \sqrt{-\frac{\Delta^2}{\cQ^2}}
\frac{\sqrt{(1 - x)^3 (1 - y)}}{x y\, (2 - x)^2}
\left( 1 - \frac{\Delta^2_{\rm min}}{\Delta^2} \right)^{3/2}
\widetilde\cT \, d \cM ,
\end{eqnarray}
where in $d^{+} \sigma_{\rightarrow}$ the proton is polarized along the 
positron beam and
\begin{equation}
\widetilde\cT = \frac{1}{2 M^2} {\rm Im}
F_2 \left( \cH^T_G + \frac{x}{2} \cE^T_G \right).
\end{equation}

Let us stress that the results (\ref{WeightCSunp}) and (\ref{WeightCSpol}) 
are valid to $1/\sqrt{\cQ^2}$ accuracy and further expansion terms from 
other structure functions can mimic the $\cos \left( 3 \phi \right)/\sin 
\left( 3 \phi \right)$ behaviour and contaminate the double helicity flip 
cross sections.

\section{Conclusion.}

In this paper we have presented weighted real photon electroproduction
cross sections which can be used as a probe for the magnitude of gluon
content in the nucleon. The opportunity to extract the real and
imaginary parts of these amplitudes is offered by the scattering of the
unpolarized lepton beam on the unpolarized and longitudinally polarized 
nucleon targets, respectively. We have given as well the formulae for NLO 
coefficient function as well as two-loop anomalous dimensions of the 
conformal operators and momentum fraction exclusive kernels thus 
completing the set of results required for study of the scaling violation 
for all twist-two SPDs at NLO.

\vspace{0.3cm}

This work was supported by DFG and BMBF (D.M.).

\end{document}